\documentclass[pre,twocolumn,superscriptaddress]{revtex4-1}

\usepackage{epsfig}
\usepackage{latexsym}
\usepackage{amsmath}
\usepackage{amssymb}
\usepackage{amsfonts}
\usepackage{graphicx}
\usepackage{wrapfig}
\usepackage[left]{lineno}
\usepackage{url}
\usepackage{graphicx}
\usepackage{dcolumn}
\usepackage{bm}
\usepackage{psfrag}

\usepackage[normalem]{ulem}  
\usepackage{color} 

\definecolor{darkgreen}{rgb}{0.0, 0.4, 0.0}
\definecolor{Mycolor1}{rgb}{1.0,0.7,0}

\begin{document}


\title{Particle flow rate in silos under rotational shear} 

\author{D. Hern\'{a}ndez-Delfin} 
\affiliation{Departamento de F\'{\i}sica y Matem\'{a}tica Aplicada, 
Universidad de Navarra, Navarra, Spain}
\author{T. Pong\'o} 
\affiliation{Departamento de F\'{\i}sica y Matem\'{a}tica Aplicada, 
Universidad de Navarra, Navarra, Spain}
\author{K. To}
\affiliation{Institute of Physics, Academia Sinica, Taipei, Taiwan R.O.C.}
\author{T. B\"orzs\"onyi} 
\email{borzsonyi.tamas@wigner.hu}
\affiliation{Institute for Solid State Physics and Optics,  Wigner Research Centre for Physics \\ P.O. Box 49, H-1525 Budapest, Hungary}
\author{R.C. Hidalgo}
\email{raulcruz@unav.es}
\affiliation{Departamento de F\'{\i}sica y Matem\'{a}tica Aplicada, 
Universidad de Navarra, Navarra, Spain}

\date{\today}

\begin{abstract}
Very recently, To et al.~have experimentally explored granular flow in a cylindrical silo, with a bottom wall 
that rotates horizontally with respect to the lateral wall \cite{Kiwing2019}. 
Here, we numerically reproduce their experimental findings, in particular, the peculiar behavior of the mass flow rate $Q$ as a function of the frequency of rotation $f$. 
Namely, we find that for small outlet diameters $D$ the flow rate increased with $f$, while for larger $D$ a non-monotonic behavior is confirmed. 
Furthermore, using a coarse-graining technique, we compute the  macroscopic density, momentum, and the stress tensor fields. 
These results show conclusively that changes in the discharge process are directly related to changes in the flow pattern from funnel flow to mass flow. 
Moreover, by decomposing the mass flux (linear momentum field) at the orifice into two main factors: macroscopic velocity and density fields, we obtain that the non-monotonic behavior of the linear momentum is  caused by density changes rather than by changes in the macroscopic velocity. 
In addition, by analyzing the spatial distribution of the kinetic stress, we find that for small orifices increasing rotational shear enhances the mean kinetic pressure $\langle p^k \rangle$ and the system dilatancy. 
This reduces the stability of the arches, and, consequently, the volumetric flow rate increases monotonically. 
For large orifices, however, we detected that $\langle p^k \rangle$ changes non-monotonically, which might explain the non-monotonic behavior of $Q$ when varying the rotational shear. 

\end{abstract}
\maketitle


\section{Introduction}
Flows involving particulate systems are commonly found in many engineering applications and natural processes \cite{Nedderman1992, Forterre_review2008,andreotti_forterre_pouliquen_2013}. In general, granular flows are complex flows involving several time and length scales, ranging from the scale of the particle deformation to the container dimensions. In the past, significant experimental and theoretical efforts were made to understand the macroscopic response of granular media in terms of their local particle-particle interactions \cite{Nedderman1992,Forterre_review2008,andreotti_forterre_pouliquen_2013}. 

The flow of particles out of a silo is a paradigmatic example of granular flow \cite{beverloo,Mankoc2007,Koivisto2017, Fullard2019, Jandaprl2012}. Decades ago, Beverloo \cite{beverloo} proposed a non-linear phenomenological correlation between the silo discharge rate $Q$ and the outlet diameter $D$.  Namely, $Q \propto (D-kd)^{5/2}$ where $d$ is the grain diameter and $k$ is a fitting parameter. This formulation rests on the assumption that the velocity of the grains scales with the outlet diameter as $\sqrt{D}$, and it uses an effective size $D-kd$.  For sufficiently large outlets, when the discharge is continuous, the correlation $Q \propto D^{5/2}$ has been tested extensively. However, for smaller outlet sizes, the flow becomes intermittent, and the system clogs randomly, and consequently, Beverloo's correlation fails to predict the flow rate values. 

More recently, researchers validated an alternative formulation, which also covers the region of small orifices where clogs frequently occur \cite{Jandaprl2012}. It accounts for the dilatancy of the system, which significantly increases with decreasing orifice diameter $D$ such that 
\begin{equation}
Q = C  \left(1-\alpha_1 e^{-D/\alpha_2} \right) D^{5/2}
\label{diego_formula}
\end{equation}
where the constant $C$ depends on the grain diameter and the curvature of the density and velocity profiles at the orifice.  
Note that in Eq.~(\ref{diego_formula}),  the exponential correction accounts for the dilatancy of the flow in relation to the aperture size.
Thus, it mimics the dilatancy dependency with the orifice diameter $D$, using 
an exponential saturation to the value $\phi_\infty$, which corresponds to the limit of big orifices, and 
$\alpha_1$ and $\alpha_2$ are fitting parameters.

In this framework, there is an interesting theoretical question, whether the clogging probability becomes zero 
above a well-defined critical orifice size, or it decreases exponentially with increasing $D$ 
\cite{To_critical,iker_critical,Thomas2015}. Practically, in systems with orifice size smaller than 
$D^* \approx 5d$, the formation of arches causes flow fluctuations, and the system will eventually clog. 
Moreover, it is known that introducing vibrations significantly changes the stability of the arches 
\cite{janda08a,Mankoc_vibration}, as well as the distribution of unclogging times 
\cite{Bulbul_Arches,Guerrero_vibration}.
However, in determining the macroscopic flow rate, vibrations play a very non-trivial role.   
Years ago, it was experimentally observed that horizontal vibrations tend to enhance the flow rate, whereas 
vertical vibrations tend to decrease it, as a function of the vibration velocity \cite{Hunt1999, Wassgren2002}. 
Although, very recently, Pascot et al.~found a non-monotonic behavior of the flow rate in a quasi-2D silo under 
vertical vibrations, depending on the vibration amplitude \cite{Pascot2020}, passing from a regime where the flow 
rate diminishes at low amplitudes to another regime where the flow rate increases. 

Very recently, To et al.~explored the discharge of a cylindrical silo with a rotating bottom \cite{Kiwing2019}. 
Interestingly, they found continuous flow for orifice sizes, notably smaller than $D^*$.  Previously, a 
numerical study of a similar system resulted in an increasing discharge rate with increasing shear rate, 
which was quantified by the Froude number $Fr = L^2 w^2/g$, where $w$ is the rotation speed and $L$  a characteristic length scale \cite{Hilton2010}. Despite considerable research effort examining clogged and non-clogged states in silo flow, a well-founded theory to successfully explain this complex response is still lacking.
   
When investigating granular flows, the researchers face several experimental restrictions, and very often, it is not possible to address 3D system behavior with all the needed details. 
In this framework, Discrete Element Modeling (DEM) is a proven alternative to examine granular systems under different boundary conditions \cite{poschel05a}. Numerically, DEM treats each particle of a granular system individually, accounting for the interaction between neighboring particles, which depends on the particle shape, friction, and elasticity.  Thus, DEM provides the macroscopic response of granular media under specific boundary conditions \cite{Weinhart2012, Weinhart2015,saraprl2015,Rubio-Largo2016,Talbot2018, Pugnaloni2020}, and all the system micro-mechanical details are crucial to understanding these responses.         

The continuous description of granular flows \cite{Staron2014, Kamrin2015, Zhou2019} is another approach which is often an efficient tool when dealing with industrial and engineering applications. The DEM data, {\it i.e.}, velocity, position, and contacts also allow building continuum fields, using coarse-grained average techniques \cite{Goldhirsch2010, Babic1997, Weinhart2013,Richard2015,Artoni2019}.
As a result, continuum fields of momentum, density, and stresses are derived. Importantly, these coarse-grained fields satisfy the mass and momentum balance equations exactly at any given time. Moreover, they are extremely useful 
for identifying relevant length and time scales \cite{Weinhart2012,Weinhart2015}, as well as, other macroscopic changes like detecting shear bands \cite{shear_bandsCG}, particle segregation \cite{segregation_CG}, and other dynamic transitions \cite{saraprl2015,Rubio-Largo2016}.
 
In this work, we numerically analyzed the granular flow in a silo with a rotating bottom. This system had been explored experimentally very recently \cite{Kiwing2019}, and motivated the numerical and theoretical analysis presented here. The paper is organized as follows: in Sec.~\ref{model}, we explain the DEM algorithm and the coarse-grained formulation \cite{Goldhirsch2010}. In Sec.~\ref{results}, the numerical results are presented and discussed in detail, shedding light on the system micro-mechanics and its relation with the system macroscopic response under this specific boundary condition.   

\begin{figure}
\begin{center}
\includegraphics[height=0.25\textheight]{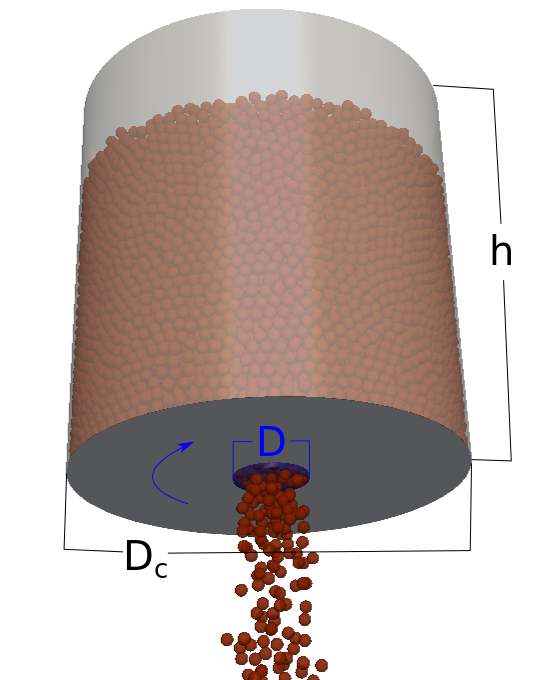}
\caption{Sketch of the numerical system, which resembles the experimental setup described in Ref.~\cite{Kiwing2019}. $D_c = 2R_c$ is the diameter of  the container.} 
\label{fig:numerical}
\end{center}
\end{figure}

\section{Numerical Model}
\label{model}

Fig.~\ref{fig:numerical} illustrates the simulated system, which resembles the experimental setup described in Ref.~\cite{Kiwing2019}. The system consists of a cylindrical container of height $h=40$ cm and radius $R_c= 9.5$ cm, with a circular aperture at the bottom wall and particles with $d=5.8$ mm. The novelty introduced in Ref.~\cite{Kiwing2019} was that the bottom of the silo could rotate about the axis of the silo, while the cylindrical wall was at rest. As in the experiment \cite{Kiwing2019}, we perform a systematic study, varying the frequency of rotation of the bottom wall from $f=0.0$ to $f=1.0$ ($\Delta f = 0.1$) all in Hz, and the radius of the orifice $R$. 

We use a discrete element modeling (DEM) implementation, consisting a hybrid $CPU/GPU$ algorithm, which allows the evaluation of the dynamics of several hundred thousand particles \cite{Owens2008,Rubio-Largo2016}. For each particle $i=1...N$, the $DEM$ algorithm solves the three translational degrees of freedom, and the rotational movement is described by a quaternion formalism. The interaction force between particle $i$ and particle $j$ reads,
\begin{equation}\nonumber
	{\vec F}_{ij} = (k_n \delta_n + \gamma_n v_r^{n})\times{\hat n} + (k_t \xi + \gamma_t v_r^{t})\times{\hat t}.
	\end{equation}
\noindent
Here we use a Hertz-Mindlin model \cite{poschel05a}, and
\begin{equation}\nonumber
k_n = \frac{4}{3}Y \sqrt{R_e \delta_n} \;\; \;\;\;\; \;\;  k_t= 8 G\sqrt{R_e \delta_n} 
\end{equation}
where the parameters $R_e$, $Y$ and $G$ are the equivalent radius, Young modulus and shear modulus, respectively. Moreover, the normal and tangential dissipation factors can be calculated as,    
\begin{equation}\nonumber
\gamma_n= 2\sqrt{\frac{5}{6}} \beta \sqrt{S_n m^*}  \;\; \;\;\;\; \;\; \gamma_t=2\sqrt{\frac{5}{6}} \beta \sqrt{S_t m^*}, 
\end{equation}
where $S_n= 2Y\sqrt{R_e \delta_n}$, $S_t=8G\sqrt{R_e \delta_n}$, $m^*=\frac{m_i+m_j}{m_i m_j}$ and $\beta=\frac{\rm ln(e_n)}{\sqrt{\rm ln^2(e_n) + \pi^2}}$. The parameter $e_n$ is the normal restitution coefficient of the particles. The tangential relative displacement ${\vec \xi}$ is kept orthogonal to the normal vector and it is truncated as necessary to satisfy the Coulomb constraint $|{\vec F}_{ij}^{t}|\leq \mu |{\vec F}_{ij}^{n}|$,  where $\mu$ is the friction coefficient. Finally, ${\vec \tau}_{ij} = {\vec b}_{ij} \times  {\vec F}_{ij}$ 
accounts for the torque corresponding to each contacting force. 
Here, ${\vec b}_{ij}$ is the brach vector from the center of particle-$i$ to the contact point between particle-$i$ and particle-$j$.

The translational equations of motion of each particle are integrated using a Verlet Velocity algorithm \cite{Verlet1968}, and a Fincham's leap-frog algorithm is used for the rotational ones \cite{Fincham1992}.  In all the simulations presented here, the system is composed of $N=30144$ particles and the contact parameters correspond approximately to particles with Young's modulus $Y = 3.0 $~GPa~~($G=Y/30$), density $\rho_{\rm p}=2655\;{\rm kg}/{\rm m}^3$, normal restitution coefficient $e_{\rm n}=0.9$ and friction $\mu=0.5$. The particle-wall interaction is modeled using the same collision parameters used for particle-particle interaction. The integration time step is set to $\Delta t = 1.0\times 10^{-6}\,s$ \cite{Antypov2011}, and all the other parameters are chosen to match the experimental conditions of Ref.~\cite{Kiwing2019}.

\subsection{Coarse-graining procedure}
\label{sec:CG}

When focusing on the macroscopic properties of granular flow, we need to obtain continuous fields from the microscopic details. For this, we use a coarse-graining method \cite{Goldhirsch2010,Weinhart2013,Richard2015,Artoni2019}, which is a well-known micro-macro mapping technique. From the positions $\vec{r}_i(t)$ and velocities $\vec{v}_i(t)$ of the particles at time $t$ in the numerical simulation, according  to~\cite{Goldhirsch2010,Weinhart2013,Richard2015,Artoni2019}, the microscopic mass density of a granular flow, $\rho(\vec{r},t)$ is defined by
\begin{equation}
\rho\left(\vec{r},t\right) = \sum_{i=1}^{N} m_i \phi\left(\vec{r}-\vec{r}_i(t)\right)
\end{equation}
\noindent where the sum runs over all the particles within the system and $\phi\left(\vec{r}-\vec{r}_i(t)\right)$
is an integrable coarse-graining function.
Similarly, the coarse-grained momentum density function $P(\vec{r},t)$ is defined by
\begin{equation}
{\vec P}(\vec{r},t) = \sum_{i=1}^{N} m_i \vec{v}_{i}(t) \phi\left(\vec{r}-\vec{r}_i(t)\right).
\end{equation}
The macroscopic velocity field $\vec{V}(\vec{r},t)$ is then defined as the ratio of momentum and density fields,
\begin{equation}
\label{velocidad}
{\vec V}(\vec{r},t) = {\vec P}(\vec{r},t)/\rho(\vec{r},t).
\end{equation}

To define the mean stress field, we use a very elegant and mathematically consistent definition of mean stress ${\sigma}_{\alpha \beta}$ introduced by Goldhirsch~\cite{Goldhirsch2010, Babic1997}. 
Following his approach, the total stress field ${\sigma}_{\alpha \beta}$ is composed of a kinetic stress field ${\sigma^k}_{\alpha \beta}$ and a contact stress field  ${\sigma^c}_{\alpha \beta}$ defined as follows. The mean contact stress tensor  is
\begin{eqnarray}
\sigma^c_{\alpha \beta} =
 -\frac{1}{2}\sum_{i=1}^{N} \sum_{j=1}^{Nc_i} f_{ij \alpha} r_{ij \beta} \int_0^1  \phi(\vec{r} - \vec{r}_i + s \vec{r}_
{ij} ) ds
\label{contact_stress}
\end{eqnarray}
where the sum runs over all the contacting particles $i,j$, whose center of mass are at
$\vec{r}_i$  and  $\vec{r}_j$, respectively. Moreover, $\vec{f}_{ij}$ accounts for  the force
exerted by particle $j$ on particle $i$ and $\vec{r}_{ij} \equiv \vec{r}_i - \vec{r}_j$.

Similarly, the mean kinetic stress field is
\begin{eqnarray}
\sigma^k_{\alpha \beta} = -\sum_i^{N} m_i v_{i \alpha}' v'_{i \beta} \phi\left(\vec{r}-\vec{r}_i(t)\right),
\label{kinetic_stress}
\end{eqnarray}
where $\vec{v}'_i$ is the fluctuation of the velocity of particle $i$, with respect to the macroscopic velocity field.
\begin{equation}
\vec{v}_{i}' (t,\vec{r}) =\vec{v}_{i} (t) - \vec{V} (\vec{r},t).
\end{equation}

Based on the previous theoretical framework, we implement a post-processing tool, which allows us to examine all the micro-mechanical properties of the particulate flow. 

\begin{figure}
	\centering
	\includegraphics[width=0.45\textwidth]{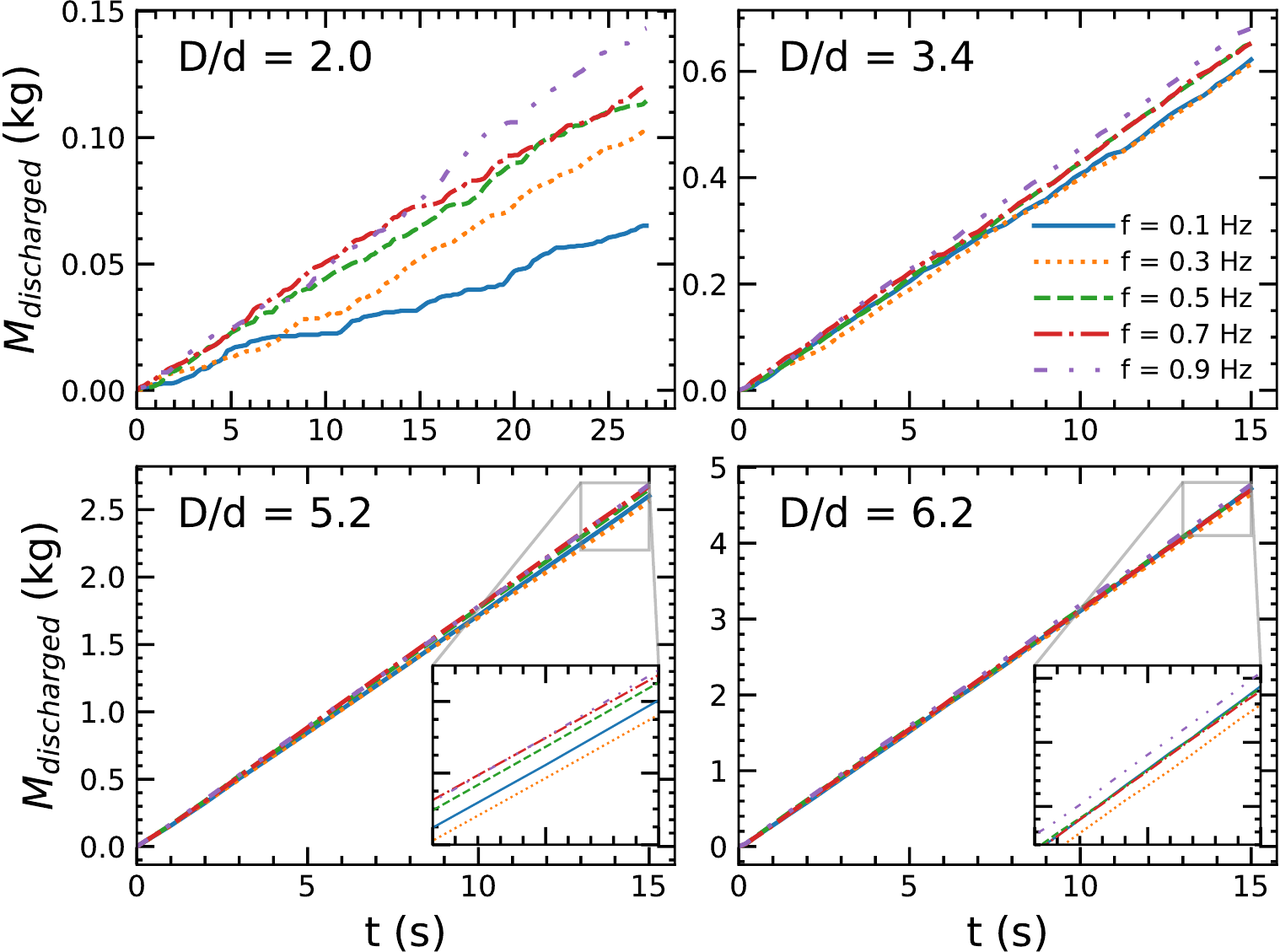}
	\caption{Discharged mass $M(t)$ versus time, obtained for various values of the orifice diameter $D$, and frequency of rotation $f$.}
	\label{fig:mas_vs_time}
\end{figure}

\section{Results and Discussion}
\label{results}
\subsection{Particle flow rate behavior}

\begin{figure}[!h]
	\centering
	\includegraphics[width=0.45\textwidth]{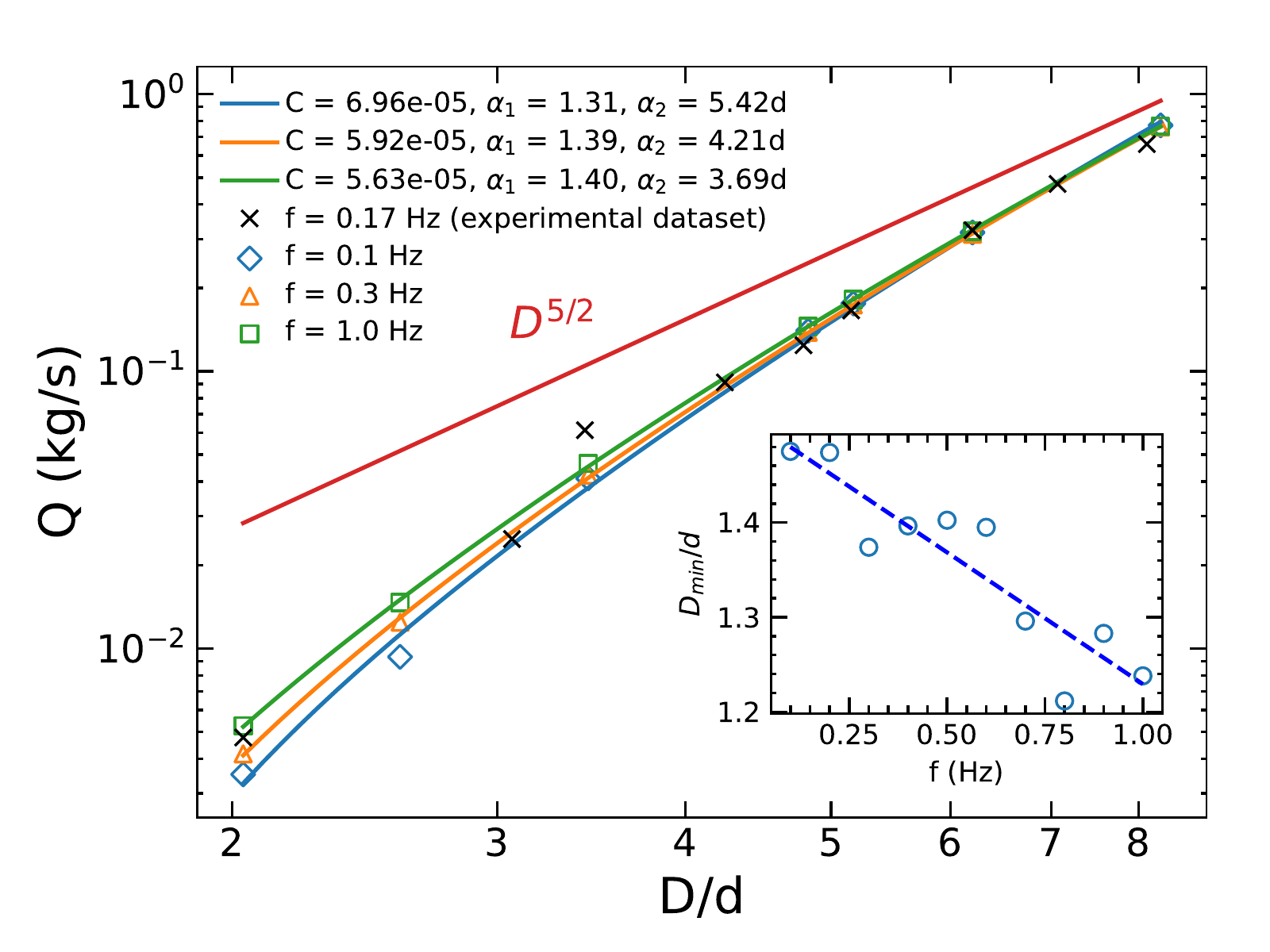}
	\caption{Mean flow rate $Q$ versus exit diameter $D$, obtained for three different values of the frequency of rotation. Lines are best fits using Eq.~(\ref{diego_formula}) as proposed in Ref.~\cite{Jandaprl2012}. The cross symbols represent the experimental data from Ref.~\cite{Kiwing2019}.}
	\label{fig:Janda_fit}
\end{figure}

As a starting point, we explore the impact of the frequency of rotation $f$ on the macroscopic 
response of the system, namely, the particle flow rate $Q$. Aiming for this objective, we carried out a systematic study while varying the orifice size $D$ and $f$. Fig.~\ref{fig:mas_vs_time} 
illustrates the discharged mass $M(t)$ versus time, obtained for various values of $D$ and $f$. 
Note that the shear perturbation introduced by the motion of the bottom wall leads to a continuous particle flow, even for orifices as small as $D = 11.8$ mm $\approx 2$ in terms of the particle diameter. 
As expected, the flow fluctuation decreases when the size of the orifice increases. 
Thus, in all cases, we can identify continuous flowing intervals where the discharged mass $M(t)$ increases 
linearly with time. It is important to mention that in static conditions (i.e., $f=0$), at $D$ lower than approximately $4.5 \times d$, the flow is quickly interrupted by the formation of stable particle  arches and permanent clogs appear.

Fig.~\ref{fig:Janda_fit} shows the flow rate $Q$ as a function of the exit size $D$ for three
values of the rotational frequency $f=0.1;\;0.3;\;1.0$ in Hz. The particle flow rate 
rises non-linearly as the orifice size increases, approaching the expected limit of Beverloo correlation $D^{\frac{5}{2}}$ \cite{beverloo}. 
This tendency was well described by Eq.~(\ref{diego_formula}) of the phenomenological model introduced in Ref.~\cite{Jandaprl2012}. The best fits according to Eq.~(\ref{diego_formula}) are presented by the continuous lines in Fig.~\ref{fig:Janda_fit}.
In all cases, we find that the fitting parameter $C$  is practically constant within our numerical uncertainties, regardless of the changes in rotation frequency. 
However, both $\alpha_1$ and $\alpha_2$ change monotonically with $f$, suggesting that the magnitude of the shear perturbation influences the system dilatancy in the region around the orifice. 

Interestingly, Eq.~(\ref{diego_formula}) implies zero mass flow rate at orifice size $D_{min}=\alpha_2 \ln{\alpha_1}$. 
Using the obtained  $\alpha_1$  and $\alpha_2$ sets of values from  the fittings, the inset of Fig.~\ref{fig:Janda_fit} shows the plot of $D_{min}$ vs. $f$ dependence. In the explored regime, the data seems to fall on a decaying straight line when increasing the rotational frequency. However, one should expect $D_{min}$ to approach nonlinearly to the physical limit $D_{min} = d$, when $f$ increases indefinitely.    
The value of $D_{min}(f)$ can be interpreted as the transition orifice diameter that separates the intermittent flow regime to the permanent clogged one, and $\alpha_1$ and $\alpha_2$ carry information of the clogged-intermittent flow transition. Although we focused our attention in large enough orifice sizes, that guaranteed continuous flow conditions; the extrapolation of $D_{min}(f=0)$ resulted approximately $1.52 \times d$, which is in good agreement with previous experimental findings \cite{Mankoc2007}.    

\begin{figure}[!t]
	\centering
	\includegraphics[width=0.45\textwidth]{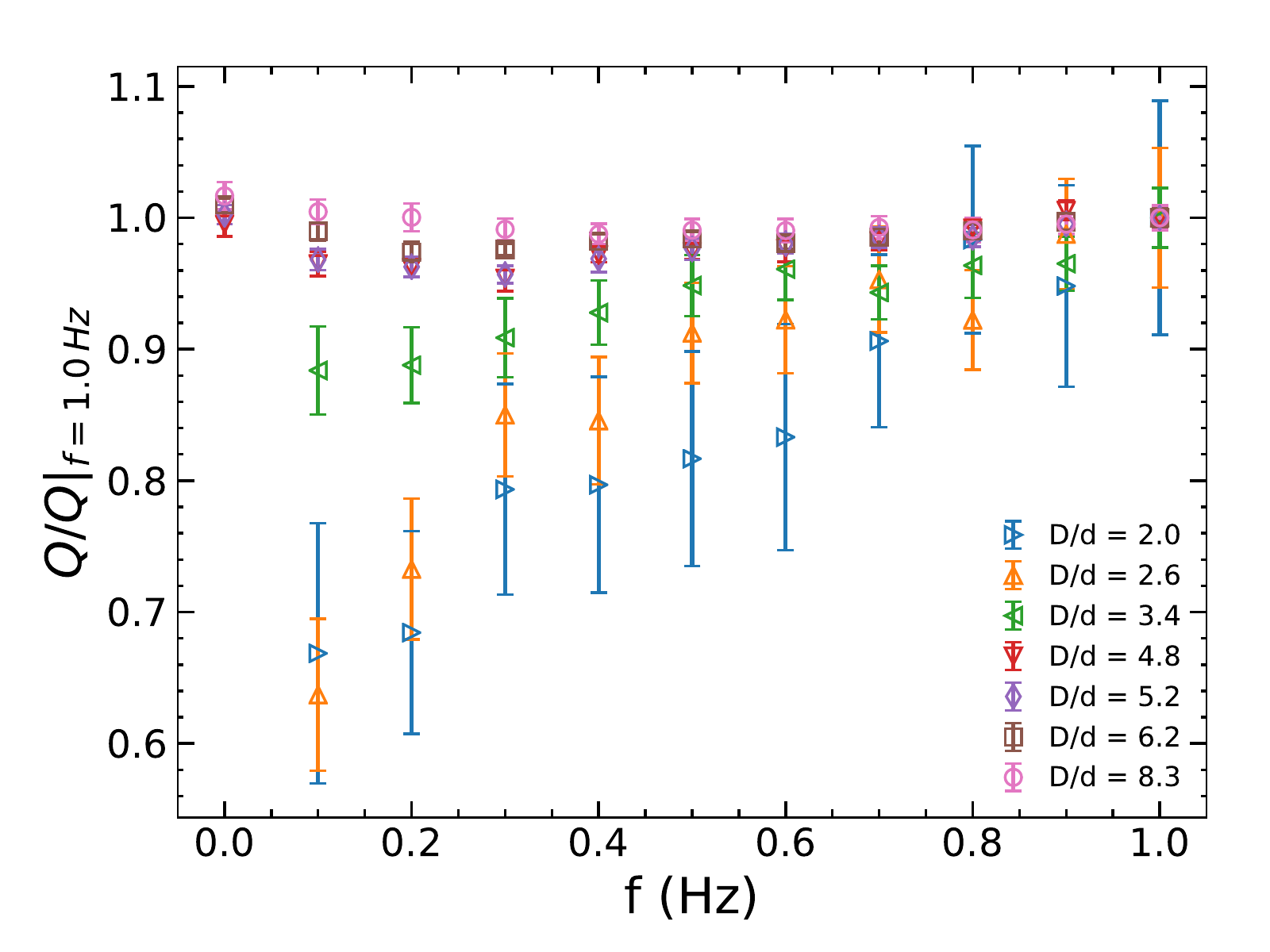}
	\caption{Normalized mean flow rate versus frequency of rotation $f$, obtained for various exit diameters $D$.
		The error-bars represent $95\%$ confidential intervals. In the case $f = 0$, permanent clogs develop when $D/d < 4.8$.} 
	\label{fig:flow_rate_all}
\end{figure}
\begin{figure}
	\centering
	\includegraphics[width=0.5\textwidth]{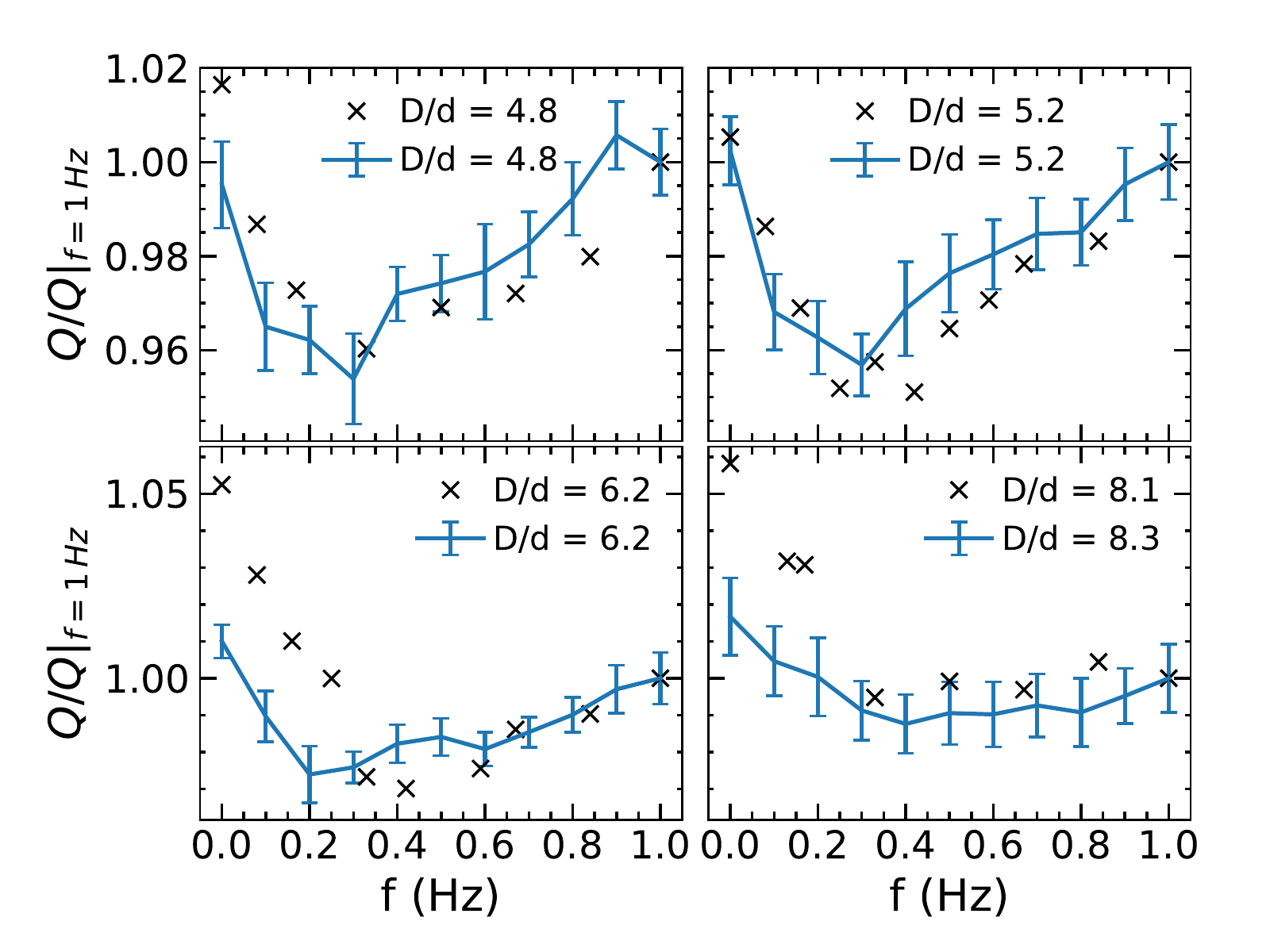}
	\caption{Normalized flow rate versus rotational frequency $f$ for a silo with exit diameter $D = 28$ mm (a), $30$ mm (b), 36 mm (c), and $48$ mm (d). In each case, the error bars represent confidence intervals for the mean with $95$ \% of confidence level. In all cases, the cross symbols represent the experimental data from Ref.~\cite{Kiwing2019}.} 
	\label{fig:flow_rate_minimum}
\end{figure}

Fig.~\ref{fig:flow_rate_all} shows the variation of flow rate $Q$ with respect to $f$, obtained for various orifice sizes. 
It is important to mention that the data of $Q$ are time-averaged values computed during flowing intervals. 
This becomes relevant for small orifice sizes ($D\leq 3.3\;d$) when flow rate fluctuations are significant.    
For convenience, the flow rate values are re-scaled with the value $Q_{f=1}$, which corresponds to $f=1.0$ Hz. 
Intriguingly, depending on the size of the aperture $D$, two distinct behaviors emerge. When $D \leq 3.3\;d$, the 
flow rate $Q$ is a strongly increasing function of $f$. For $D>3.3\;d$, however, $Q$ changes smoothly with $f$,  denoting a weakly non-monotonic behavior. Taking a closer look at the second regime, Fig.~\ref{fig:flow_rate_minimum} illustrates the data obtained for large orifices, focusing on the specific data range. Even though the changes are of the order of $5\%$ of $Q_{f=1}$, the existence of a minimum is obvious, denoting a change in the discharge process, {\it i.e.}, the flow rate decreases for low rotation speeds starting from $f = 0$,  then at a certain value of $f$, it starts to increase. Very recently, Kiwing To and coworkers \cite{Kiwing2019} found this trend experimentally. Remarkably, our numerical procedure reproduced those outcomes quantitatively with high accuracy.

The two most frequent flow patterns in silos and bins are the {\it funnel flow} and {\it mass flow}. Thus, when the stress profile along the silo is not smooth enough to ensure sliding along its walls, a {\it funnel flow} develops. Consequently, particles flow through a channel at the silo center and a stagnant zone develops close to the walls. In {\it mass flow}, however, the stress profile is smooth enough to ensure the flow of all the particles within the system. 
In Ref.~\cite{Kiwing2019}, the authors speculated that the change in the discharge process with rotational shear
might be related to a crossover in the flow pattern, from a funnel flow to mass flow. 
While their arguments were based only on visual inspection of the top surface of their experimental system, we can directly observe the change in the flow pattern from the macroscopic fields measured in our numerical simulations.

\begin{figure}
	\includegraphics[height=0.3\textheight]{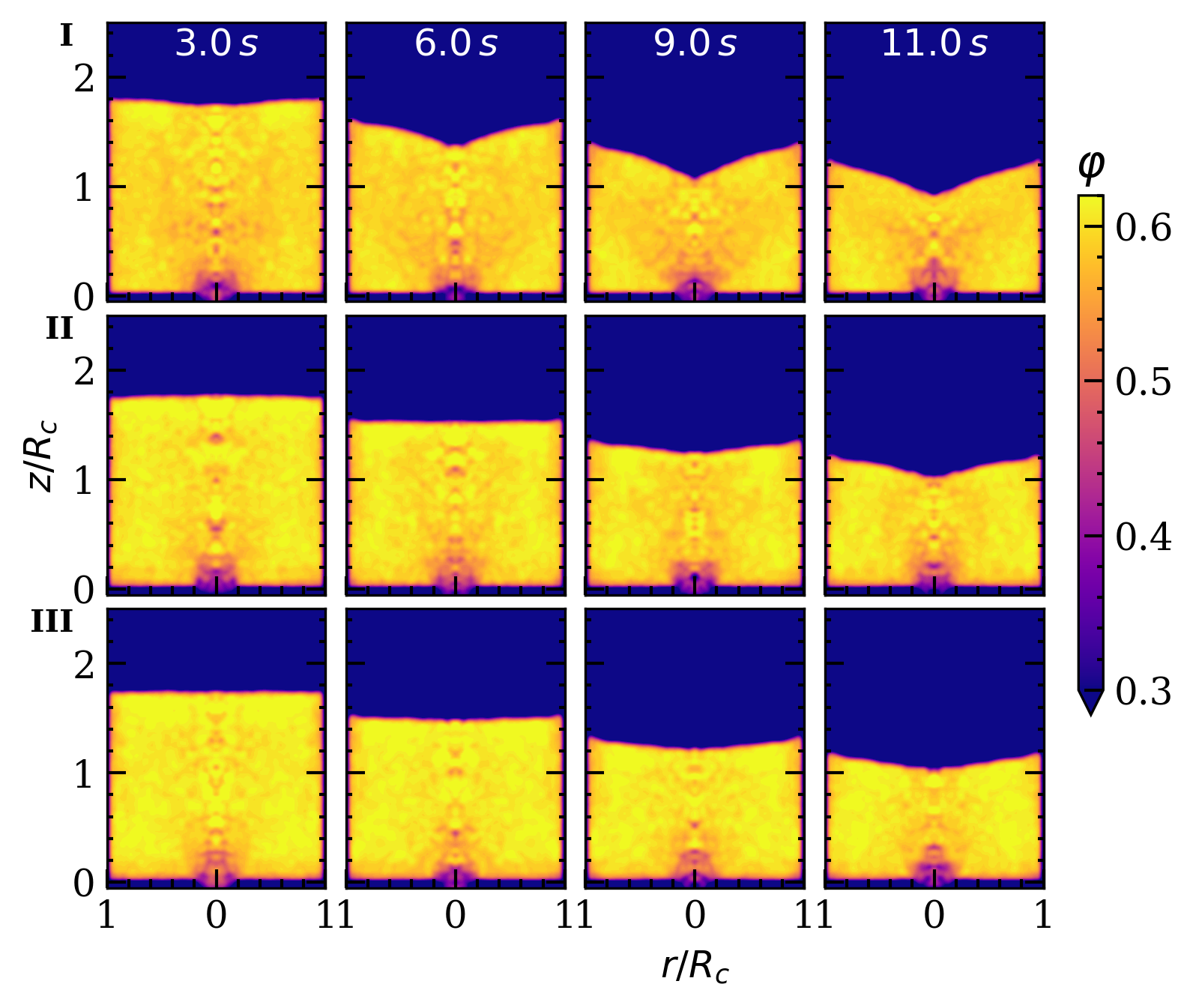}
	\caption{
		Color map representing the packing fraction spatial profiles $\varphi\left(r,z,t \right)$ obtained for $D = 36$ mm $(6.20\;D/d)$ and various rotational frequencies, in row {\bf I} ($f=0$\;Hz), in row {\bf II} ($f=0.3$\; Hz), and in row {\bf III} ($f=1.0$\;Hz). The corresponding time is indicated on the top panel. In computation, we use a truncated Gaussian coarse-graining function $\phi(\vec{r})$ with a coarse-grained scale equals to the particle radius.
	} 
	\label{fig:density_maps}
\end{figure}

\subsection{Continuous field view}
The numerical simulations allow accessing the micro-mechanical details of the granular flow, both inside the silo and at the orifice. 
Using the DEM data of each individual particle, we computed the macroscopic fields: volume fraction $\varphi\left(\vec{r},t\right) = \rho\left(\vec{r},t\right)/\rho_{\rm p}$, macroscopic velocity $\vec{V}(\vec{r},t)$  and kinetic stress ${\bf \sigma}^k(\vec{r},t)$. Taking advantage of the cylindrical symmetry, we average the vertical and radial components of the studied quantities within an azimuthal representative volume element of uniform size. As a consequence, the macroscopic fields result in cylindrical coordinates $r$ and $z$ in units of the radius $R_c$ of the cylindrical silo. 

\begin{figure}
	\includegraphics[height=0.3\textheight]{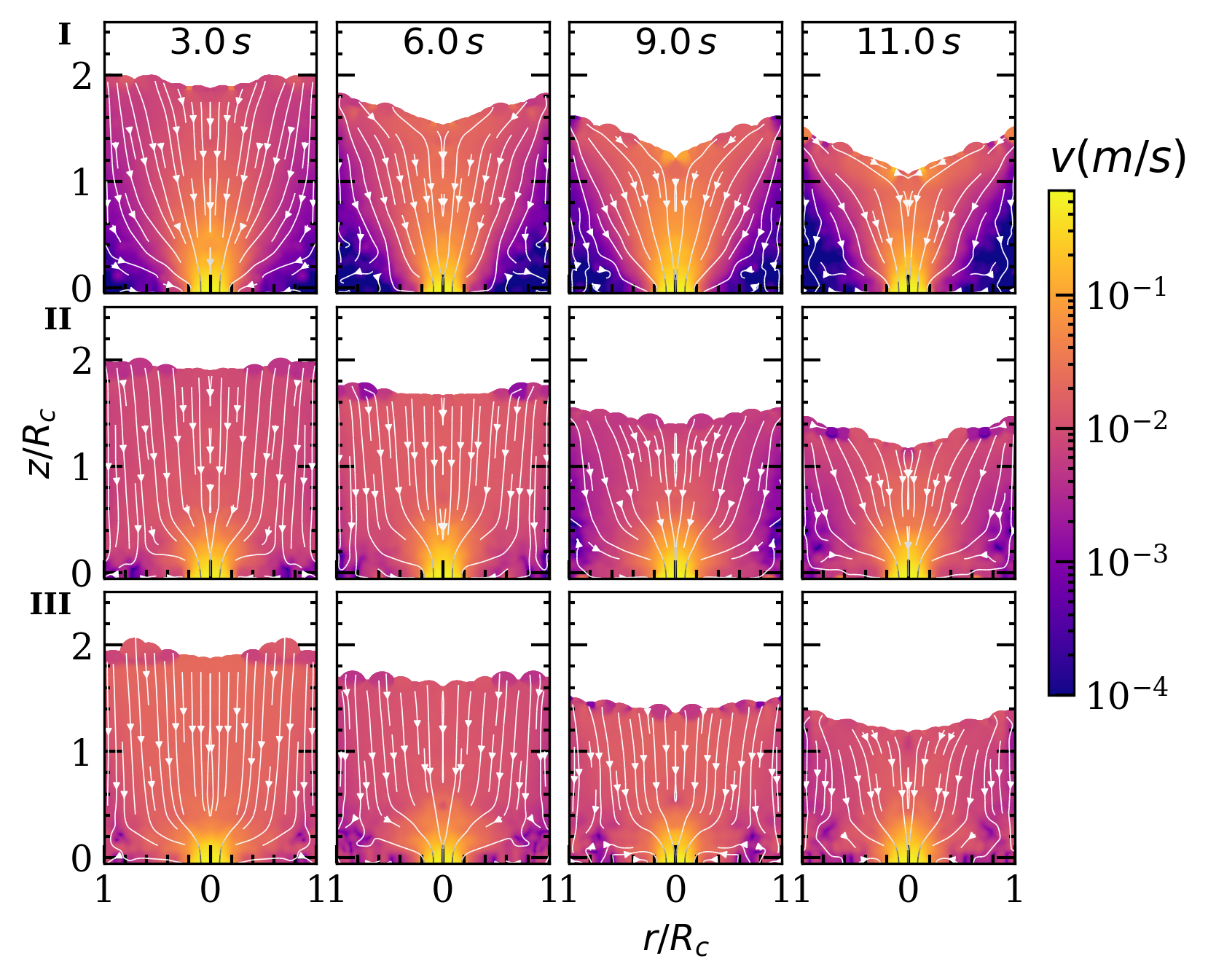}
	\caption{
		Color map representing the velocity field $v\left(r,z,t \right)$  obtained for $D = 36$ mm $(6.20\;D/d)$ and various rotational frequencies, in row {\bf I} ($f=0$\;Hz), in row {\bf II} ($f=0.3$\; Hz), and in row {\bf III} ($f=1.0$\;Hz). 
		The streamlines are also illustrated. The corresponding time is indicated on the top panel. In computation, we use a truncated Gaussian coarse-graining function $\phi(\vec{r})$ with a coarse-grained scale equals to the particle radius.}
	\label{fig:velocity_maps}
\end{figure}

\begin{figure}
	\includegraphics[height=0.3\textheight]{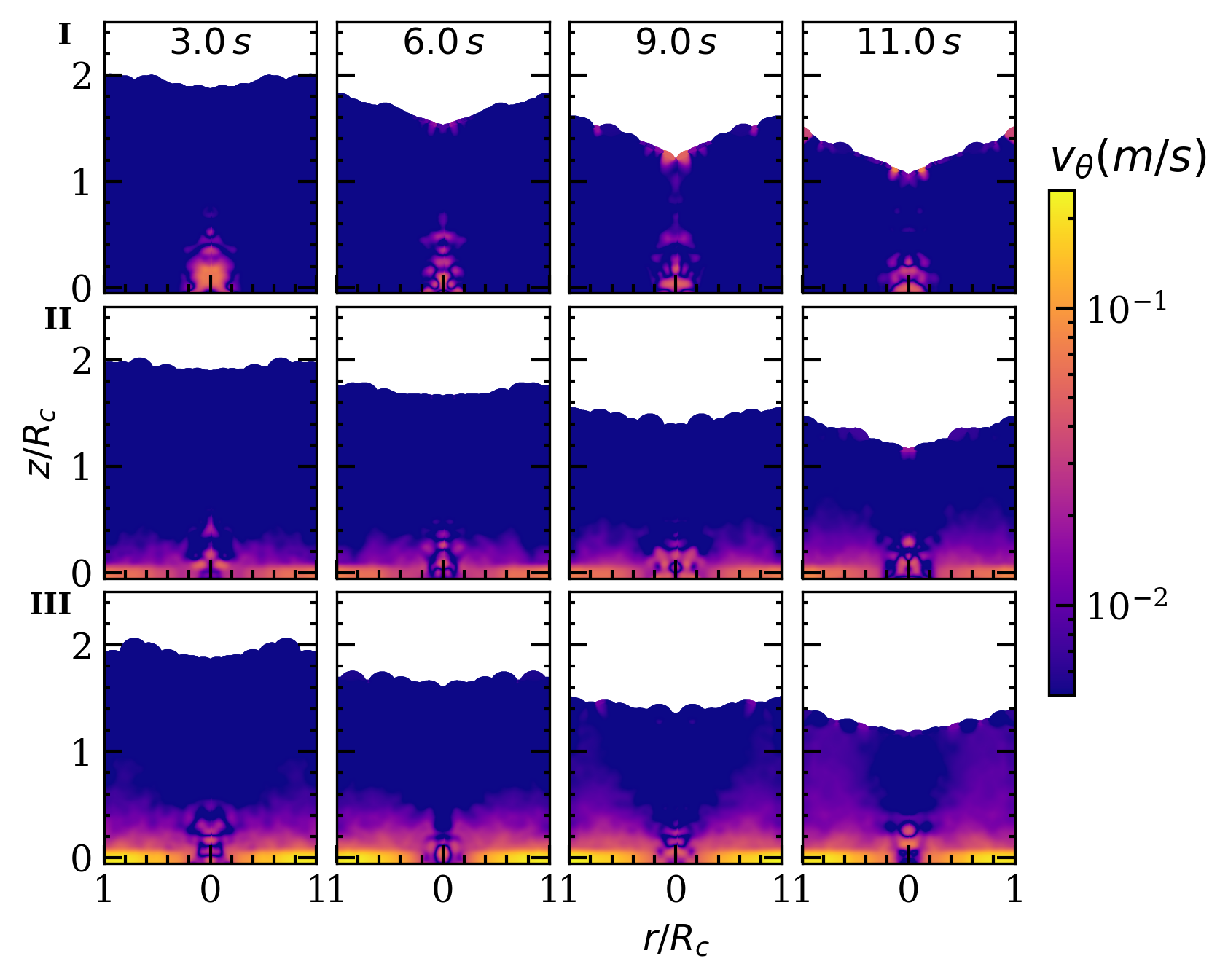}
	\caption{
		Color map representing the azimuthal velocity field $v_{\theta}\left(r,z,t \right)$  obtained for $D = 36$ mm $(6.20\;D/d)$ and various rotational frequencies, in row {\bf I} ($f=0$\;Hz), in row {\bf II} ($f=0.3$\; Hz), and in row {\bf III} ($f=1.0$\;Hz). 
		The corresponding time is indicated on the top panel. In computation, we use a truncated Gaussian coarse-graining function $\phi(\vec{r})$ with a coarse-grained scale equals to the particle radius.}
	\label{fig:tangential_velocity_maps}
\end{figure}

Taking advantage of a detailed continuum description, we clarify the nature of the change in the discharge process, {\it i.e.}, the flow rate decreasing for low rotation speeds, followed by an enhancement for high rotation speeds.
Fig.~\ref{fig:density_maps} illustrates the volume fraction fields $\varphi\left(r,z,t \right)$ as color maps, covering the entire system. 
The data is displayed in three rows, which correspond to three different rotational frequencies $[f = 0 \,;\,0.3;\,1.0]$ in Hz, respectively. 
Moreover, the fields allow visualizing the time evolution, and each column corresponds to a specific time $[t =3;6;9;11]$ in seconds.
Note that in the static case {($f = 0$)  funnel flow develops: the particles mainly flow through the central core of the silo (see also supplementary material). As a result, the volume 
fraction field $\varphi\left(\rho,z,t \right)$ is heterogeneous, a shear band develops, and a stagnant 
region is observed close to the lateral wall. 
Moreover, right from the beginning of the process, a depression appears at the center of the top surface, and its size increases as the silo empties. 

\begin{figure}
	\centering
	\includegraphics[width=0.40\textwidth]{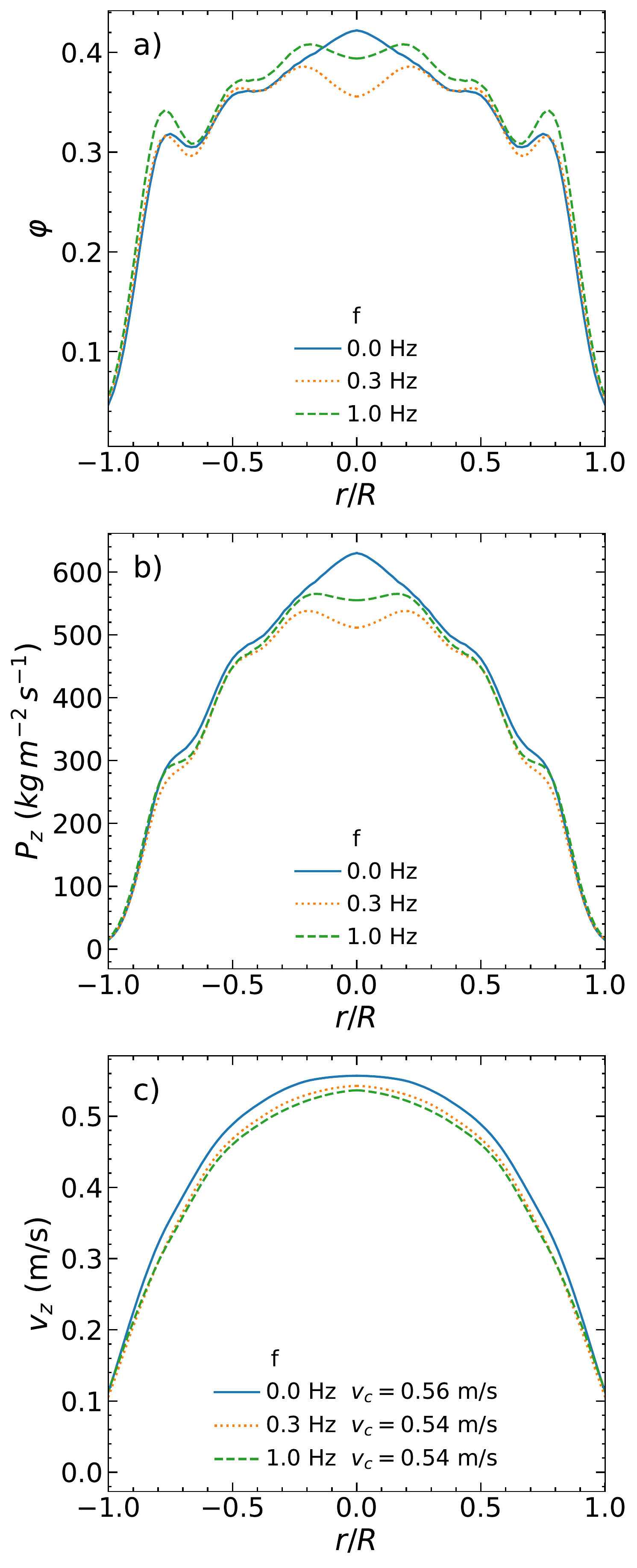}
	\caption{Spatial profiles at the orifice ($z=0$) for $D = 36$ mm $(6.20\;D/d)$, in a) average density field $\langle \varphi\left( r  \right) \rangle$, b) average momentum on the vertical direction $\langle P_z\left( r \right) \rangle$ and c) average velocity on the vertical direction  $\langle V_z\left( r \right) \rangle$.  In each case, finding corresponding to rotational frequency $[f = 0;\,0.3;\,1.0]$ in Hz are shown.} 
	\label{fig:d_p_v}
\end{figure}

\begin{figure*}
	\includegraphics[height=0.15\textheight]{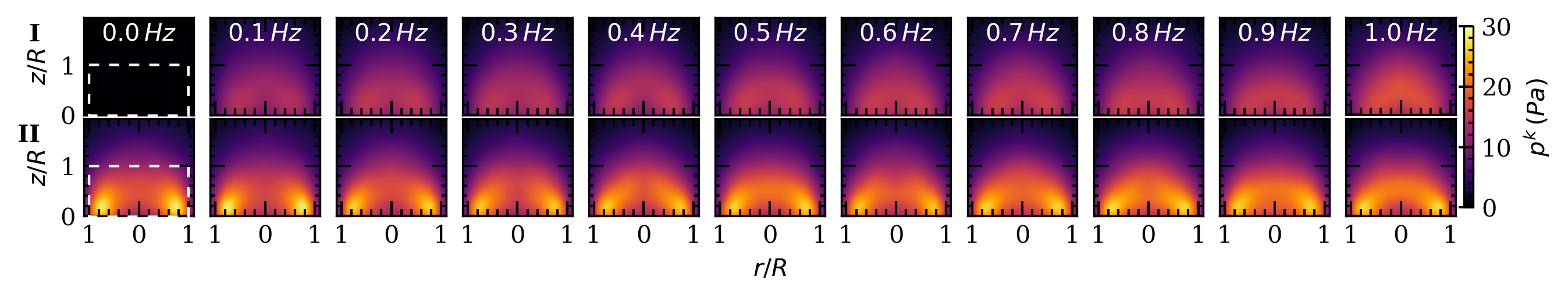}
	\caption{Spatial profiles of kinetic pressure $\langle p^k\left( r,z \right) \rangle=Tr( \langle \sigma^k\left( r,z \right) \rangle_t)$, time-averaged during flowing states. The figures display 
		data obtained for two sizes of the orifice, row {\bf I}: $D=20$ mm $(3.45\;D/d)$,  and row 
		{\bf II}: $D = 36$ mm $(6.20\;D/d)$.  The corresponding rotational frequency is indicated on the top panel. In computation, we use a truncated Gaussian coarse-graining 
		function $\phi(\vec{r})$ with a coarse-grained scale equals to the particle radius.} 
	\label{fig:kinetic_fields1}
\end{figure*}

When the bottom wall rotates (Fig.~\ref{fig:density_maps}, rows {\bf II} and {\bf III}), however, particles located close to the base are mobilized. 
Consequently, the rotational shear perturbs the system even at large distances from the bottom, reordering events 
concatenate, and the system fluidizes significantly. 
As a  result, no stagnant region forms, the top surface remains flat, and the appearance of the depression is notably delayed. All of these are signatures of mass flow behavior. Pascot et al.~\cite{Pascot2020} found similar behavior in a quasi-2D silo under vertical vibrations, at low vibration amplitude, an increase of vibrations reduces the size of the stagnant zones, and consequently, the flow rate decreases as well. 

The differences in the bulk flow patterns are more evidenced by the spatial features of the velocity field 
${\vec V}\left(r,z,t \right)$. Fig.~\ref{fig:velocity_maps} displays   
the streamlines of ${\vec V} \left(r,z,t \right)$, while the colors represent the 
magnitude of the speed. 
When the bottom of the silo is not moving (row {\bf I}), the velocity field is rather heterogeneous, and strong velocity gradients emerge in both radial and vertical directions. Besides, the streamlines are considerably curved, drawing a complex flow pattern over the whole system. Note that at the center of the silo, the magnitude of the speed $v(r,z,t)$ is significantly larger, in comparison with the region close to walls (stagnant zone), and $v(r,z,t)$ rises notably in the region of the orifice. 

On the other hand, the movement of the bottom wall perturbs the systems dynamics significantly, resulting in pronounced changes in the velocity field (Fig.~\ref{fig:velocity_maps}, rows {\bf II} and {\bf III}).
It induces smooth enough conditions, which ensure that particles in the whole container can move downwards. 
We found that the down-up collisional energy transmission reduces the strength of the velocity gradients, in both radial (not shown) and vertical directions. Thus, as the rotational speed increases, the perturbation impacts higher locations, where the velocity gradient in the radial direction practically diminishes.

Fig.~\ref{fig:tangential_velocity_maps} displays the azimuthal velocity $v_{\theta}$ to complement the results presented in Fig.~\ref{fig:velocity_maps} and reaffirm the fact that the stagnant zone is mobilized when $f$ is high enough. Comparing the rows, one can see that the impact of the rotational shear affects not only the radial dependency of the tangential velocity but also its dependency with the height.

Let us focus on the region near the orifice, where we perform a quantitative analysis of the macroscopic fields, examining their relation with the resulting particle flow rate $Q$. 
First, we compute the macroscopic solid fraction $\varphi\left( r,z,t \right)$, momentum ${\vec P} \left(r,z, t\right)$ and velocity ${\vec V} \left( r,z,t \right)$ fields, at the cross-section of the orifice, located at $z=0$. Assuming that the system reaches a steady state such that the time average of the fields are well defined, namely,  $\langle \varphi\left( r  \right) \rangle$, $\langle {\vec P}\left( r\right) \rangle$ and $\langle {\vec V}\left( r \right) \rangle$. 
Fig.~\ref{fig:d_p_v} displays the average density field $\langle \varphi\left( r  \right) \rangle$, the average vertical momentum $\langle P_z\left( r \right) \rangle$ and the average vertical velocity $\langle V_z\left( r \right) \rangle$  for three values of the rotational frequency $[f = 0 \,;\,0.3;\,1.0]$ in Hz. Interestingly, the vertical momentum  (Fig.~\ref{fig:d_p_v}b) has a weak but noticeable non-monotonic behavior when changing $f$, i.e. the values of $\langle P_z\left( r \right) \rangle$ at the orifice are larger for $f=0$ and 
$f=1.0$ Hz, compared to the case of $f=0.3$ Hz. Besides, the same applies to the density profiles 
$\langle \varphi\left( r  \right) \rangle$. The velocity profiles however change less, and the change is monotonic with rotation speed.
 
Mass conservation requires that the particle flow rate crossing the section of the 
orifice is $Q= \int_S {\vec P} \cdot d{\vec A} = \int_S \rho  \cdot {\vec V} \cdot d{\vec A}$. 
Thus, the data for $\langle P_z\left( r \right) \rangle$ (Fig.~\ref{fig:d_p_v}b) at the orifice are consistent with the non-monotonic behavior of the $Q$ vs. $f$ curves calculated from particle data (Fig.~\ref{fig:flow_rate_minimum}), and those obtained experimentally (Fig.~7 of Ref.~\cite{Kiwing2019}). Stepping forward, our numerical data suggest that the non-monotonic behavior of the momentum is rather caused by solid-fraction changes than by macroscopic velocity changes. Thus, the micro-mechanical analysis clearly indicates that the shear perturbation created by the rotating wall induces a nontrivial system dilatancy in the region of the orifice, and consequently, a non-monotonic behavior of the flow rate when changing the rotation speed. 

In granular flows, the kinetic stress, which is the stress associated with velocity fluctuations, can be used to identify relevant length and time scales as well as dynamic transitions \cite{saraprl2015,Rubio-Largo2016}.
Fig.~\ref{fig:kinetic_fields1} displays color maps that represent the spatial profiles of kinetic pressure $\langle p^k\left( r,z \right) \rangle$, which is defined as the trace of the kinetic stress tensor Eq.~(\ref{kinetic_stress}), namely, $\langle p^k\left( r,z \right) \rangle=Tr( \langle \sigma^k\left( r,z \right) \rangle_t)$. 
For clarity, Fig.~\ref{fig:kinetic_fields1} illustrates data for two sizes of the orifice, row {\bf I}:  $D=20$ mm $= 3.45\;D/d$, and  row {\bf II}: $D= 36$ mm $= 6.20\;D/d$, at rotational frequency from $f = 0.0$ Hz to $f=1.0$ Hz. 
When computing the fields, we use a truncated Gaussian coarse-graining function $\phi(\vec{r})$ with a coarse-grained scale equals to the particle radius, and the color maps represent the time-averaged values computed during flowing intervals. 
  In general, we find that the values of kinetic pressure are more significant in the region of the orifice, 
and are diminishing with height. This suggests that the mass transport in the silo is mainly advective. 
However, as the particles get closer to the exit, their individual movements decorrelate from the 
global flow. Both for static conditions or for systems with rotating bottom, a region resembling a {\it free fall arch} is observed, where the kinetic pressure is maximum. 
After crossing this region, the particles fall mainly driven by gravity. Interestingly, in row II one can see a slightly non-monotonic change in the color map intensity, when increasing the frequency.

\begin{figure}
	\includegraphics[height=0.3\textheight]{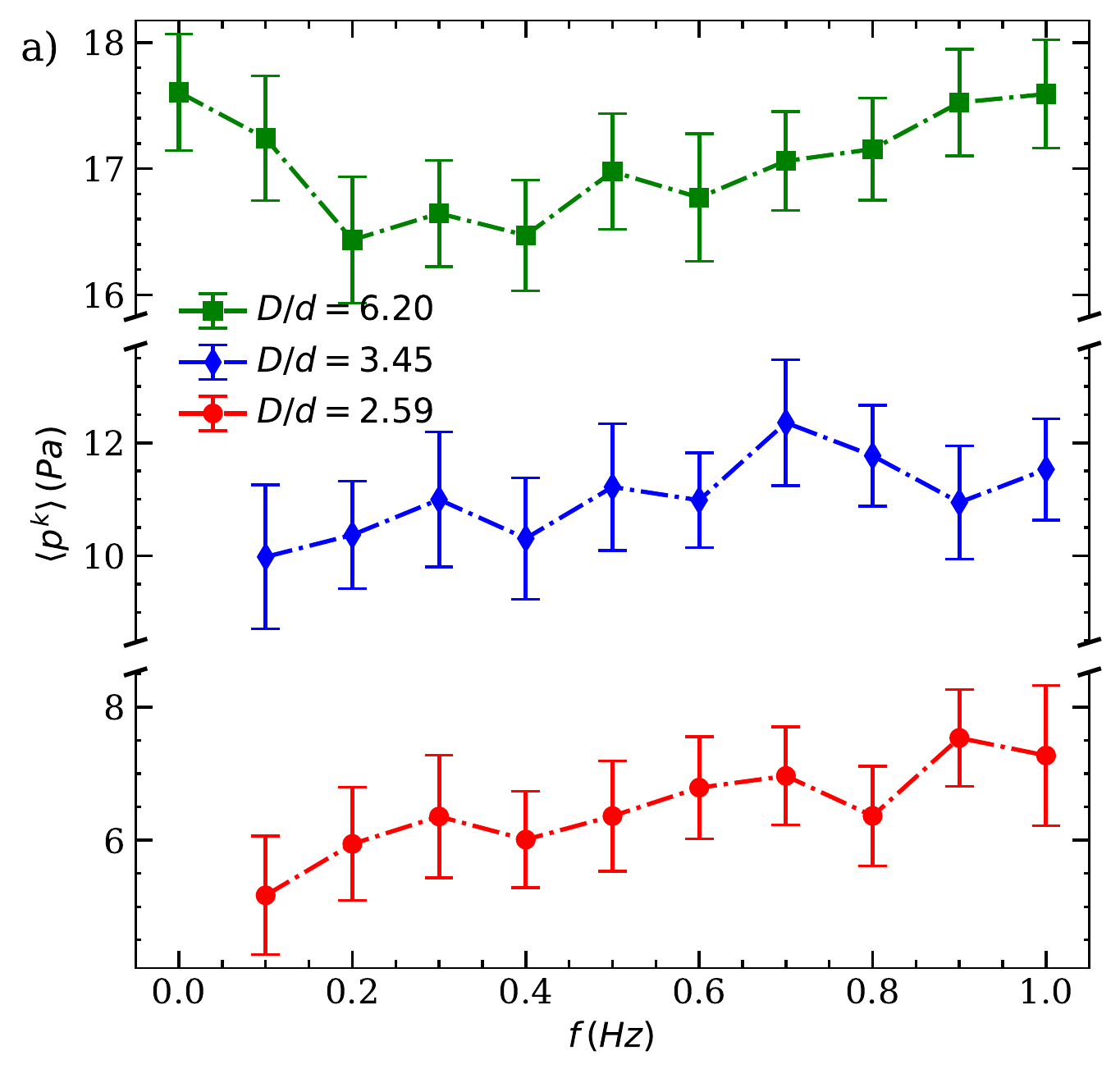}
	\includegraphics[height=0.3\textheight]{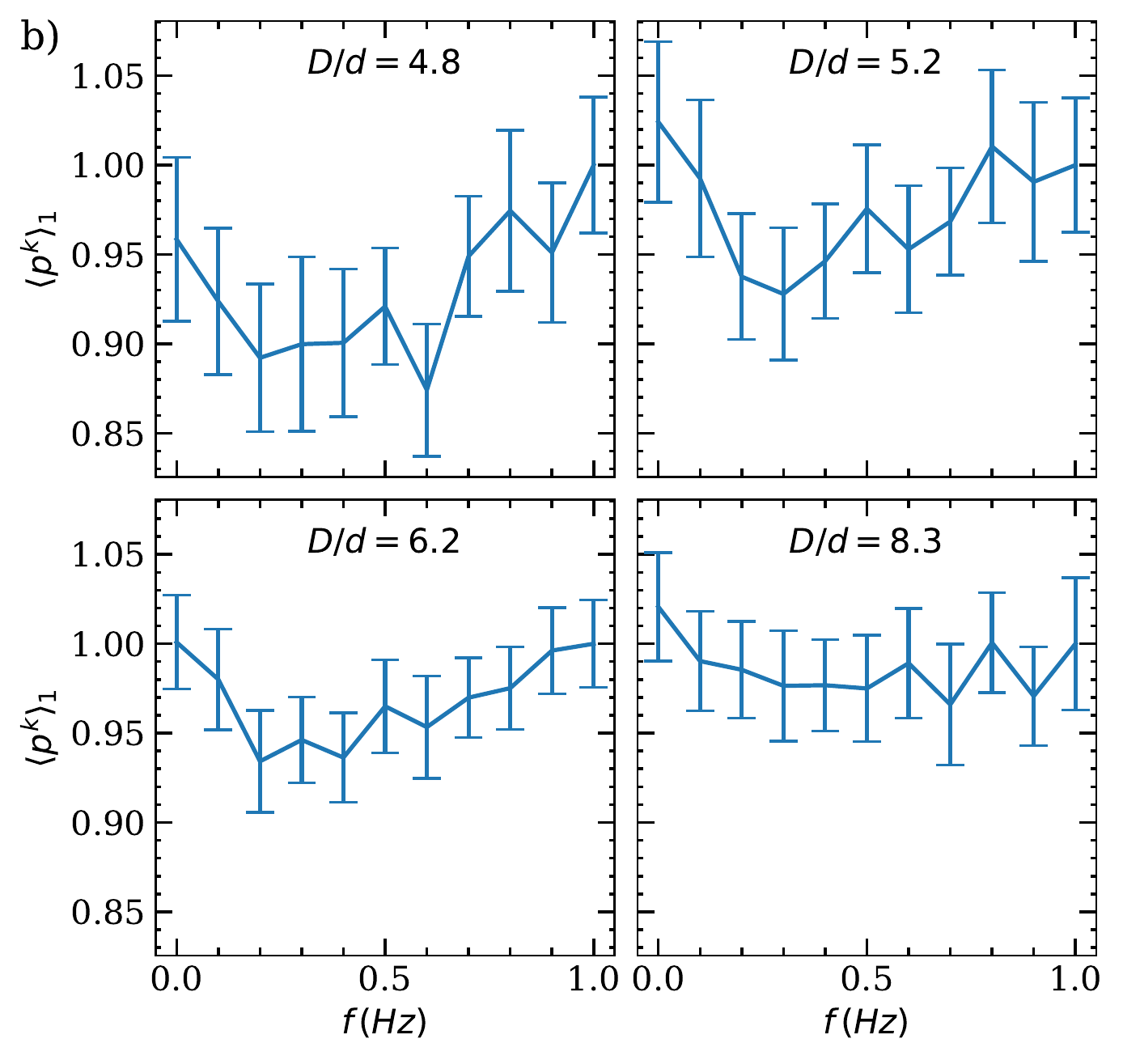}
	\caption{Mean kinetic pressure $\langle p^k \rangle=Tr(\langle\sigma^k\rangle_{ts})$, averaged in the region of the orifice as a function of the rotation frequency $f$. a) Data obtained for $D=15$ mm,  $D=20$ mm and $D =36$ mm. b) Data $\langle p^k \rangle_1$=($\langle p^k \rangle$ normalized by its value at $f=1$ Hz) for $D=28$ mm,  $D=30$ mm, $D =36$ mm and $D = 48$ mm. 
		In each case, the error bars represent confidence intervals for the mean with $95$ \% of confidence level.}
	\label{fig:kinetic_vs_w}
\end{figure}

In order to better quantify the effect of rotation speed on the stress associated with velocity fluctuations, we calculated the mean kinetic pressure $\langle p^k \rangle=Tr(\langle\sigma^k\rangle_{ts})$ in the region of the orifice. This was done by averaging the mean kinetic pressure in a cylindrical region centered at the orifice with a height of $\delta h = R$ and a radius of $R$. Here, $R=D/2$ is the radius of the orifice (see dashed rectangles in Fig.~\ref{fig:kinetic_fields1}).
As we see in Fig.~\ref{fig:kinetic_vs_w}a, for small orifices, the mean kinetic pressure increases monotonically with increasing rotation speed. 
Presumably, the rotational shear induces a monotonically increasing dilatancy, which reduces the stability of the arches. As a result, the volumetric flow rate also increases monotonically.  
For large orifices (see Fig.~\ref{fig:kinetic_vs_w}b), however, we observe that the kinetic pressure $\langle p^k \rangle$ changes non-monotonically. It drops to a minimum value (about 95\% of its value at $f=1$ Hz) at some intermediate values of $f$.  In this range, the region with maximum kinetic pressure gets slightly more diffused than at small or large values of $f$ (see Fig.~\ref{fig:kinetic_fields1}). Apparently, this non-monotonic trend in the kinetic pressure is connected to changes in the discharge process, a flow rate decreasing for low rotation speed, whereas flow rate enhancement for high rotation speed \cite{Kiwing2019}. 

{\it Summarizing}, we reported DEM simulations and coarse-graining analysis, which 
reproduced a granular flow quantitatively in a cylindrical silo, with a bottom wall that rotates horizontally with respect to the lateral wall \cite{Kiwing2019}. We find that depending on the size of the aperture $D$, two distinct behaviors emerge. For small orifices, the flow rate $Q$  results in a strongly increasing function of the rotational frequency $f$. For large $D$, however, $Q$ changes smoothly with $f$, denoting a slightly non-monotonic behavior.
Stepping forward, our findings shed light on the nature of the flow when changing the rotational frequency, and prove that changes in the discharge process are directly related to changes in the flow pattern, from funnel flow to mass flow, with increasing $f$. 
We also observe that the momentum profiles at the orifice present a non-monotonic behavior when changing $f$. Remarkably, these findings are consistent with the non-monotonic behavior of the flow rate obtained from particle data numerically, and in laboratory experiments \cite{Kiwing2019}. Additionally, a close examination of the density and velocity profiles indicates that the non-monotonic behavior of the momentum is caused by the change in density instead of the changes in macroscopic velocity. Examining the profiles of kinetic stress, for small orifices, we show that the rotational shear induces a monotonically increasing kinetic pressure $p^k$ and dilatancy. This seems to reduce the stability of arches, and, as a result, the volumetric flow rate monotonically increases as well. For large orifices, however, we detected that the mean kinetic pressure $\langle p^k \rangle$ changes non-monotonically, which explains the non-monotonic behavior of $Q$ with the strength of the rotational shear. 

\section*{Acknowledgments} 
This project has received funding from the European Union's Horizon 2020 research and innovation programme under the Marie Sklodowska-Curie grant agreement CALIPER No 812638, by the Spanish MINECO (FIS2017-84631-P MINECO/AEI/FEDER, UE Projects) and by the NKFIH (Grant No. OTKA K 116036). D.H. acknowledges Asociaci\'on de Amigos de la Universidad de Navarra.


\end{document}